\documentclass[pra,twocolumn,showpacs]{revtex4}

\usepackage{graphicx}

\begin{document}

\title{Single atoms in a standing-wave dipole trap}

\date{\today}

\author{Wolfgang Alt}
\email{w.alt@iap.uni-bonn.de}
\author{Dominik Schrader}
\author{Stefan Kuhr}
\author{Martin M\"uller}
\author{Victor Gomer}
\author{Dieter Meschede}
\affiliation{Institut f\"ur Angewandte Physik, Universit\"at Bonn,
Wegelerstr. 8, D-53115 Bonn, Germany}

\begin{abstract}
We trap a single cesium atom in a standing-wave optical dipole
trap. Special experimental procedures, designed to work with
single atoms, are used to measure the oscillation frequency and
the atomic energy distribution in the dipole trap. These methods
rely on unambiguously detecting presence or loss of the atom using
its resonance fluorescence in the magneto-optical trap.
\end{abstract}


\maketitle

\section{Introduction}

In the last decade, optical dipole traps have become a standard
tool for trapping ultracold samples of neutral atoms (see
\cite{Grimm 00,Balykin 00} and references therein). In
far-off-resonance traps \cite{Miller 93} atoms are trapped in a
nearly conservative potential, where they exhibit a low
spontaneous scattering rate leading to long coherence times up to
several seconds \cite{Davidson 95}. These features, in combination
with a great variety of possible trap designs and the ability to
create time dependent trapping potentials, allow the study of
classical and quantum chaos \cite{Milner 01}, production and
manipulation of Bose-Einstein condensates \cite{Stamper 98} and
investigations of ultracold atom mixtures \cite{Mosk 02}. These
applications require the transfer of large numbers of cold atoms
into the dipole trap \cite{Kuppens 00}.

In contrast, this work focuses on experiments with only a single
or a few trapped atoms. Our long-term objective is the controlled
manipulation of quantum states of individual atoms. On the way to
achieve this goal, we have recently demonstrated the possibility
of manipulating the position and the velocity of a single atom
with high precision using a movable standing-wave optical
potential \cite{Kuhr 01, Schrader 01}.

To fully take advantage of the available techniques, it is
essential to access all trap parameters and to understand
fundamental effects such as lifetimes and heating effects. On the
one hand, trapping of a few atoms avoids collisional loss and
heating mechanisms associated with large numbers of atoms
\cite{Kuppens 00}. On the other hand, standard observation schemes
like time-of-flight methods based on direct imaging of an atomic
cloud are not applicable.

Our methods rely on unambiguously detecting presence or loss of an
atom using its resonance fluorescence from a magneto-optical trap
(MOT)~\cite{Hu 94}. The ability to transfer an atom from the MOT
into the dipole trap and back without any loss \cite{Frese 00}
allows us to determine its survival probability after any
intermediate experimental procedure in the dipole trap. Mastering
this single-atom preparation and detection is the basis of the
results presented in this paper.

In Sec.~\ref{sec:dipole trap} we briefly describe the standing
wave dipole trap and our experimental setup. In
Sec.~\ref{sec:heating} the relevant heating mechanisms for atoms
in our trap are evaluated and put in relation with the observed
lifetime. A measurement of the energy distribution of the atoms in
the trap is presented in Sec.~\ref{sec:adiabatic}, as well as the
calculation of the adiabatic cooling involved. In
Sec.~\ref{sec:three-beam} we use the ability to manipulate the
dipole potential in various ways to determine the axial
oscillation frequency of the atoms, again using only one atom at a
time. Finally we summarize our results and point out future
possibilities.

\section{\label{sec:dipole trap}Standing-wave dipole trap}

Our dipole trap consists of two counter-propagating Gaussian laser
beams with equal intensities and parallel linear polarizations.
With their optical frequencies $\omega$ and $\omega + \Delta
\omega$ ($\Delta \omega \ll \omega$) they produce a position- and
time dependent dipole potential
\begin{equation}
 V(z,\rho,t,U_0) = U_0 \frac{ w_0^2}{w^2(z)}\ e^{
 -\frac{2\rho^2}{w^2(z)}}\cos^2 \! \! \left(\frac{\Delta \omega}{2} t
 - kz \right) .
\label{eq:potential}
\end{equation}
Here, $\lambda=c/\omega$ is the optical wavelength, $w^2(z)=w_0^2
\left(1+z^2/z_0^2\right)$ is the beam radius with waist $w_0$ and
Rayleigh length $z_0 = \pi w_0^2/\lambda$.

Both dipole trap laser beams are derived from a Nd:YAG laser
($\lambda = 1064$~nm), which is far red detuned from the Cesium
D$_1$- and D$_2$-transitions (894~nm and 852~nm). In this case the
maximum trap depth $U_0$ is given by
\begin{equation}
 U_0 = \frac{\hbar \Gamma}{2} \frac{P}{\pi w_0^2 I_0}
 \frac{\Gamma}{\Delta} ,
 \label{eq:U0}
\end{equation}
where $\Gamma = 2\pi \times 5.2$~MHz is the natural linewidth of
the Cesium D$_2$-line, $I_0 =1.1$~mW/cm$^2$ is the corresponding
saturation intensity and $P$ is the total power of both laser
beams. Note that for red detunings ($\Delta < 0$) the dipole
potential (\ref{eq:potential}) provides three-dimensional
confinement with a trap depth of $|U_0|$. For alkalis the
effective detuning $\Delta$ is given by \cite{Grimm 00}
\begin{equation}
 \frac{1}{\Delta}=\frac{1}{3}\left(\frac{1}{\Delta_1}+\frac{2}{\Delta_2}
 \right),
\end{equation}
where $\Delta_i$ is the detuning from the D$_i$-line. Here,
$\Delta=- 2\pi \times 64$~THz. The laser beam parameters are
$w_0=30~\mu$m, $z_0=2.7$~mm with a total power of $P=4$~W, which
yields a potential depth $U_0$ of 1.3~mK.


An atom of mass $m$ trapped in such a standing-wave potential
oscillates (in harmonic approximation) with frequencies
\begin{eqnarray}
 \Omega_{\text{z}} &=& 2 \pi \sqrt{\frac{2U_0}{m\lambda^2}} \\
 \Omega_{\text{rad}} &=& \sqrt{\frac{4U_0}{m w_0^2}}
\end{eqnarray}
in axial and radial directions, respectively. In our case
$\Omega_{\text{z}} / 2\pi = 380$~kHz and $\Omega_{\text{rad}} /
2\pi = 3.1$~kHz.

\begin{figure}
 \includegraphics[width=8.5cm]{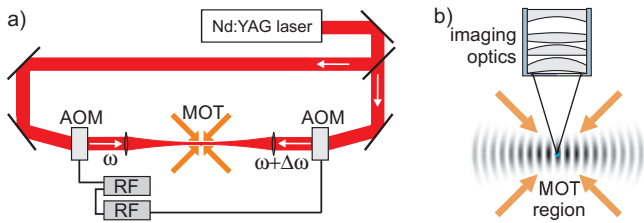}
 \caption{\label{fig:setup}Experimental setup. (a) MOT and dipole trap
 are overlapped in the center of a vacuum cell (not shown). Acousto-optical
 modulators (AOMs) are used to control the frequencies of both laser beams
 which form the dipole trap. Synchronized frequency generators (RF) supply
 the AOMs with phase-continuous frequency ramps in order to transport the
 atom.
 (b) The imaging optics collects fluorescence of the atom in the MOT.}
\end{figure}

Figure~\ref{fig:setup} shows a schematic view of the experimental
setup (see \cite{Schrader 01} for more details). A magneto-optical
trap with a high magnetic field gradient serves as a source of
single cold atoms \cite{Frese 00}. The fluorescence light from the
MOT is collected by imaging optics covering a solid angle of $0.02
\times 4 \pi$ \cite{Alt 01} and is detected by an avalanche
photodiode (APD). From each atom in the MOT we obtain up to $5
\times 10^4$ counts per second on a stray light background of only
$2 \times 10^4$~s$^{-1}$. This allows us to determine the number
of trapped atoms within 10~ms.

These atoms can be transferred from the MOT into the dipole trap
or back by operating both traps simultaneously for several 10~ms.
When the focus of the dipole trap laser is carefully superimposed
with the MOT, this transfer occurs without any loss of atoms
\cite{Frese 00,Schrader 01}.

An atom initially trapped in the stationary standing wave dipole
trap (laser beam frequency difference $\Delta \omega = 0$) can be
moved along the optical axis by changing $\Delta \omega$ which
causes the potential wells to move at the velocity
$v=\lambda\Delta\omega/4\pi$. To control the frequency difference
$\Delta \omega$, both dipole trap laser beams pass through
acousto-optical modulators (AOMs), which are set up in double-pass
configuration to avoid angular deviation of the beams. While both
AOMs are driven with the same frequency $\omega_{\rm AOM} = 2 \pi
\times 100$~MHz the standing wave pattern is at rest and atoms can
be loaded into the dipole trap. To accelerate them along the
dipole trap axis one of the AOMs is driven by a phase-continuous
linear frequency ramp. In a similar fashion they can be
decelerated and brought to a stop at a predetermined position
along the standing wave \cite{Kuhr 01,Schrader 01}.

\section{\label{sec:heating}Heating mechanisms and lifetime}

Without additional cooling, the lifetime of atoms in a dipole trap
is ultimately limited by heating. A fundamental source of heating
in dipole traps is spontaneous scattering of trap laser photons.
Due to the large detuning of the trapping laser the photon
scattering rate at the maximum trapping laser intensity is
\begin{equation}
 R_s \approx \frac{U_0 \Gamma}{\hbar \Delta}
\end{equation}
is only 14~s$^{-1}$. Each photon adds on average one recoil energy
$E_{\rm r}=(\hbar k)^2/2m$ on absorption and on spontaneous
emission. Therefore the energy $E$ of an atom in the dipole trap
potential increases as $\dot{\langle E\rangle}=2R_s
E_r$~\cite{Grimm 00}.


The above scattering rate yields a recoil heating rate of about
$\dot{\langle E \rangle}=0.9~\mu$K/s which is negligible in our
experiment. Heating due to dipole force
fluctuations~\cite{Dalibard 85} is at least four orders of
magnitude smaller than the recoil heating.

Technical heating can occur due to intensity fluctuations and
pointing instabilities of the trapping laser beams as discussed in
detail in Ref.~\cite{Gehm 98}. In the first case, fluctuations
occurring at twice the trap oscillation frequency $\Omega_0$ can
parametrically drive the oscillatory atomic motion. For a spectral
density of the relative intensity noise $S(\Omega)$ of the
trapping laser and in harmonic approximation the energy increases
exponentially according to Ref.~\cite{Gehm 98}
\begin{equation}
 \dot {\langle E \rangle} = \gamma \langle E \rangle,~\text{with}~
 \gamma = \frac{\pi \Omega_{0}^{2}}{2} S(2 \Omega_0).
\end{equation}
Even for the free-running industrial laser used here with a
relative intensity noise spectral power density of
$3\times10^{-11}/$Hz at $2*\Omega_{\rm rad}$ and
$3\times10^{-14}/$Hz at $2*\Omega_{\rm z}$ the heating time
constant is $\tau = \gamma^{-1} \approx 300$~s and 20~s,
respectively.

In the case of pointing instability, shaking of the potential at
the trap oscillation frequency increases the motional amplitude.
With $S(\Omega_0)$ being the spectral density of the position
fluctuations the heating rate is given by~\cite{Gehm 98}
\begin{equation}
 \dot {\langle E \rangle}=\frac{\pi}{2} m \Omega_0^4 S(\Omega_0) .
 \label{eq:point}
\end{equation}
In previous experiments with a running-wave dipole trap, using the
same laser but more tightly focused to $w_0 = 5~\mu$m, we have
observed lifetimes of one minute~\cite{Frese 00}. The smaller
focus leads to a much higher radial oscillation frequency
($\Omega_{\rm rad} \propto w_0^{-2}$). From the very strong
dependency (\ref{eq:point}) of the heating rate on the oscillation
frequency $\Omega_0$ we infer that the pointing instabilities in
radial direction are negligible in our current, less strongly
focused dipole trap.

\begin{figure}
 \includegraphics[width=8.5cm]{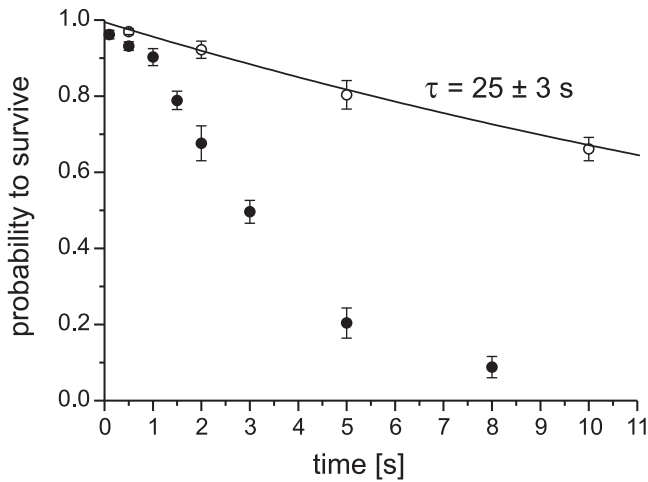}
 \caption{\label{fig:lifetime}Lifetime measurement with (filled circles)
 and without phase noise (hollow circles) at otherwise identical
 conditions.
 In the latter case the decay is purely exponential and probably due to
 background gas collisions.}
\end{figure}

All heating mechanisms described above, which are intrinsic to any
dipole trap, are not observable in this experiment and the
measured trap lifetime of 25~s is limited by background gas
collisions, see Fig.~\ref{fig:lifetime}. However, in our
experiments there is an additional technical noise due to
fluctuations of the relative phase $\Delta \phi$ between both AOM
drivers. This phase noise is directly translated by the AOMs into
position fluctuations $\epsilon$ of the dipole potential along the
standing-wave axis $\langle \epsilon^2 \rangle = \langle \Delta
\phi ^2 \rangle /k^2$. The rms phase noise amplitude
$\sqrt{\langle \Delta \phi ^2 \rangle} \approx 10^{-3}$~rad has
directly been measured by heterodyning both output signals of the
AOM drivers.

When this noise is evenly distributed over 1~MHz bandwidth and
$\Omega_0=380$~kHz, equation~(\ref{eq:point}) yields a heating
rate of 4~mK/s. At higher oscillation amplitudes the harmonic trap
approximation presumed in equation~(\ref{eq:point}) breaks down
and the oscillation frequency goes to zero which slows down the
heating process.

We used a numerical simulation to obtain a realistic estimate of
the lifetime in the anharmonic trapping
potential~\ref{eq:potential}. The one-dimensional equation of
motion in the potential $V(z,t) = U_0 \cos^2[k(z + \epsilon (t))]$
is integrated numerically, starting with the atom at rest at $z =
0$, until it leaves the potential well $|z| < \lambda /4$. The
potential is shaken with a gaussian white noise $\epsilon (t)$
with a bandwidth of 1~MHz and $\sqrt{\langle \Delta \phi ^2
\rangle} \approx 10^{-3}$~rad. This results in an average lifetime
of 2~s, in reasonable agreement with the experimental life time of
about 3~s in the presence of phase noise
(Fig.~\ref{fig:lifetime}). The different heating rates are
summarized in Table~\ref{tab:heatingrates}.

\begin{table}
 \caption{\label{tab:heatingrates}Heating mechanisms in the dipole
 trap and corresponding heating rates. For the resonant and parametric
 excitation see Section~\ref{sec:three-beam}.}
 \begin{ruledtabular}
  \begin{tabular}{lr}
   Heating effect & Heating rate \\
   \hline
   recoil heating & $9 \times 10^{-4}$~mK/s (calc) \\
   dipole force fluctuation heating & $10^{-7}$~mK/s (est) \\
   laser intensity fluctuations (radial) & $4 \times 10^{-3}$~mK/s (calc) \\
   laser intensity fluctuations (axial) & $6 \times 10^{-2}$~mK/s (calc) \\
   laser pointing stability (radial) & not observable \\
   AOM phase noise (axial) & 4~mK/s (calc) \\
   & $0.4$~mK/s (obs) \\
   resonant excitation (axial) & 10~mK/s (obs) \\
   parametric excitation (axial) & 10~mK/s (obs) \\
  \end{tabular}
 \end{ruledtabular}
\end{table}

\section{\label{sec:adiabatic}Adiabatic cooling and energy distribution}

The standard method of measuring the energy distribution of
trapped atoms is the time-of-flight technique. There, the trap is
switched off instantaneously and the velocity distribution of the
atoms in the trap is inferred from an image of their spatial
distribution after ballistic expansion. This method cannot be used
in our case because with only a single atom in the trap it would
require very many repetitions to get useful statistics.

A technique compatible with single atoms for measuring the energy
distribution in the trap is to reduce the potential depth and to
observe whether the atoms are lost. However, if this reduction of
the potential is done quickly compared to the atomic oscillation
period, the instantaneous kinetic energy determines whether the
atom escapes from the lowered potential. Thus, the loss
probability depends on the phase of the oscillation at the moment
the potential depth is reduced.

\begin{figure}
 \includegraphics[width=8.5cm]{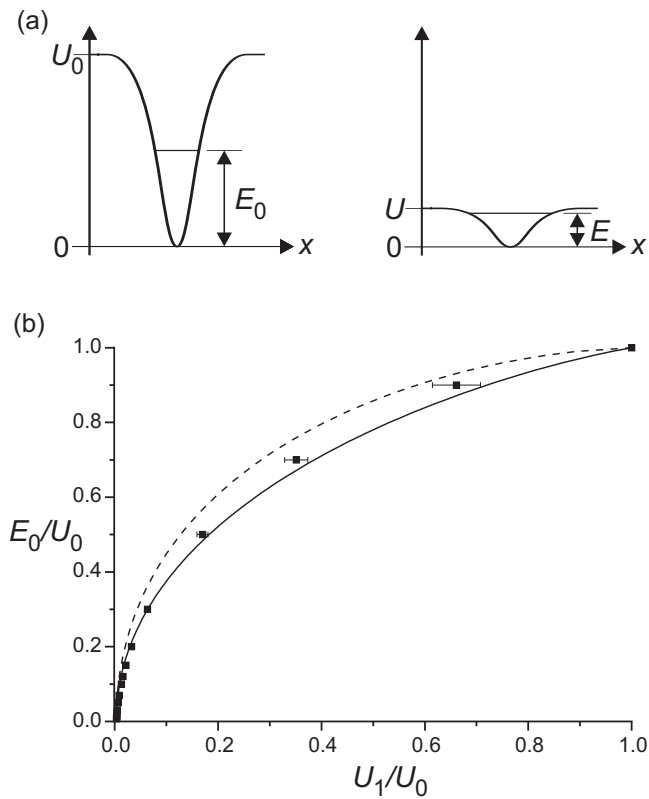}
 \caption{\label{fig:adiatheo}(a) When the trap depth is adiabatically reduced
 from $U_0$ to $U$, the energy of the atom inside the trap also decreases
 from $E_0$ to $E$.
 (b) Atoms with energy $E_0$ in the original potential of depth $U_0$
 escape when the trap depth is reduced to $U_1$.
 Solid line: one-dimensional model, axial motion, $V(x,U)=U[1-\cos^2(kx)]$;
 dashed line: radial motion, $V(x,U)=U[1-\exp(-2x^2/w_0^2)]$, squares:
 three-dimensional numerical simulation, the bars indicate the range over
 which the atoms escape.}
\end{figure}

If, in contrast, the trap depth is reduced slowly compared to the
oscillation period, i.e.~adiabatically, the trap depth $U_1$ at
which the atom escapes is a function of its total initial energy
$E_0$ only. By changing the potential depth from its initial value
$U_0$ to a value $U$, the energy of the atom is also changed from
$E_0$ to $E$, due to adiabatic cooling, see
Fig.~\ref{fig:adiatheo}(a). The atom escapes when the reduced trap
depth $U$ falls below $E$.

\subsection{\label{sec:adiabatictheory}Theory}

In a one-dimensional conservative potential $V(x, U)$ of depth
$U>0$ the action $S = \oint \! p\, dx$ remains invariant under
adiabatic variation \cite{Landau}, where the integration is
carried out over one oscillation period. If the potential is
symmetric, $V(-x, U)=V(x, U)$, the action can be written as
\begin{equation}
 S (E, U) = 4 \int_0^{x_{\text{max}}}\!\!\!\!\!\!\!\!\! dx
 \sqrt{2m [E-V(x,U)]} = \text{const.} ,
 \label{eq:inv}
\end{equation}
where $E$ is the energy of the atom and $x_{\text{max}}$ is the
turning point of the oscillatory motion given by
$V(x_{\text{max}},U)=E$.

Eq.~(\ref{eq:inv}) allows us to calculate the initial atomic
energy $E_0$ from the measured trap depth $U_1$, at which the atom
is lost. Using the invariance of $S$ we numerically solve $S(E_0,
U_0)=S(U_1, U_1)$ and show the resulting initial atomic energy
$E_0$ as a function of $U_1$ for both axial and radial motion in
Fig.~\ref{fig:adiatheo}(b).

The invariance of $S$ only holds for changes in $U$
infinitesimally slow compared to the oscillation frequency
$\Omega$, i.e.\ for $|{\dot{\Omega}}/{\Omega^2}| \to 0$. In order
to optimally lower the potential within a limited time we keep
${\dot{\Omega}}/{\Omega^2}$ constant. This requires $\Omega(t)
\propto 1/t$, which corresponds to, in harmonic approximation,
$U(t) \propto 1/t^2$. Smoothing the sudden transition from
$U(t)=U_0$ to $U(t) \propto 1/t^2$ at $t=0$ further improves the
adiabaticity. In summary, the trap depth is reduced according to
the function
\begin{equation}
 \label{eq:uvont}
 U(t) = \left\{
 \begin{array}{lll} \displaystyle
  U_0 & \text{for} & t \leq 0\\
  \displaystyle U_0 \left( 1 - \frac{t^2}{4T_{\text{c}}^2}
  \right)\vspace{0.1cm} &\text{for} & 0 < t \leq T_{\text{c}}
  \sqrt{2}\\
  \displaystyle U_0 \frac{T_{\text{c}}^2}{t^2}& \text{for} & t >
  T_{\text{c}} \sqrt{2}
\end{array}
\right.
\end{equation}
until it reaches $U_1$, with a characteristic decay time of
$T_{\text{c}} = 3$~ms. This keeps $\left|
\dot{\Omega}_{\text{rad}} / \Omega_{\text{rad}}^2 \right| < 0.02$.
A graph of $U(t)$ used in the experiment, including a waiting time
of 15~ms and a ramp up back to $U_0$, is shown in
Fig.~\ref{fig:energydist}(a). Note that due to the anharmonicity
of our potential $\Omega \to 0$ for $E \to U$, which always
violates the adiabaticity condition right before the atom leaves
the trap. However, this energy region is relatively small and the
corresponding error is in the order of $\pm2$\% of the initial
energy $E_0$.

The one-dimensional theory presented so far can only be applied to
a separable three-dimensional potential $V(x,y,z) =
V_1(x)+V_2(y)+V_3(z)$, where the equations of motion decouple. The
dipole trapping potential (\ref{eq:potential}) is not separable
and therefore effectively couples the motional degrees of freedom.
This leads to the possibility of a slow energy exchange between
them, the timescale of which can be long compared to the
oscillation period. Hence, the lowering of the potential is not
adiabatic with respect to this energy exchange time. This raises
the question whether the total atomic energy is responsible for
the escape of the atom, or rather the motional energy in the
direction of the preferred escape, i.e. along gravity.

To obtain quantitative information on the adiabatic cooling in
three dimensions, classical atomic trajectories were calculated in
a simplified time-varying potential, where $|z| < \lambda/4 \ll
z_0$ and therefore $w(z)$ has been approximated by $w_0$:
\begin{equation}
 \label{eq:simplepot}
 V(x,y,z,t) = U(t)\cos^2(kz)e^{-\frac{2(x^2+y^2)}{w_0^2}} + mgy;
\end{equation}
for $U(t)$ see eq.~(\ref{eq:uvont}) and
Fig.~\ref{fig:energydist}(a). Atoms with a fixed energy $E_0$ but
otherwise random starting coordinates are subjected to the
simulated adiabatic lowering, in order to find out at which trap
depth $U_1$, or what range of trap depths, they escape.

The algorithm for determining random starting coordinates for a
fixed initial energy $E_0$ first randomly distributes $E_0$ onto
the three energies $E_x,E_y,E_z$. It then chooses random phases
for the oscillations in the three directions, to divide each of
these energies into a potential and a kinetic fraction. These are
used to calculate starting coordinates and velocities.

The equations of motion in potential (\ref{eq:simplepot}) are
solved numerically, and atoms which depart more than $3w_0$ from
the origin are counted as lost. For given values of the initial
energy $E_0$ and minimal potential depth $U_1$ up to 120
trajectories are calculated to estimate the survival probability
for the atoms with a statistical error of $\pm0.05$ . Then $U_1$
is varied to find the value where the survival probability equals
$0.5$, see Fig.~\ref{fig:adiatheo}(b). Additionally the $1 \sigma$
range of trap depths, over which the survival probability drops
from 0.84 to 0.16, is shown as error bars. The three-dimensional
simulations of the adiabatic cooling process agree qualitatively
with the one-dimensional model. Due to the imperfect adiabaticity
of the chosen $U(t)$ (\ref{eq:uvont}) atoms of one energy $E_0$ do
not escape at exactly one trap depth $U_1$, but over a range of
about $\pm 10$\% of $U_1$. This could be improved by making the
lowering of the potential even slower.

\subsection{\label{sec:adiabaticmeasurement}Measurement of the
 energy distribution}

\begin{figure}
 \includegraphics[width=8.5cm]{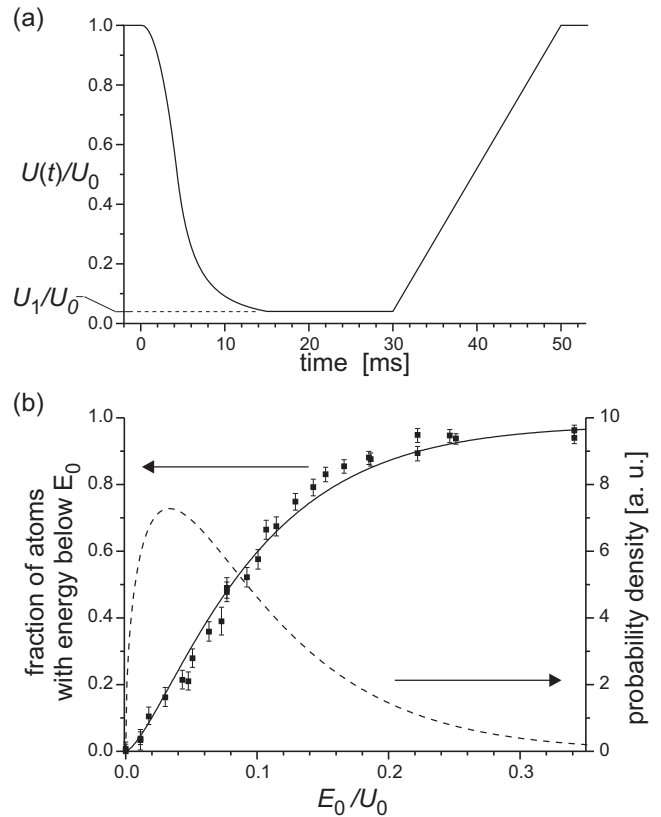}
 \caption{\label{fig:energydist}
 (a) Temporal variation of the potential depth for measurement of the
 energy distribution: Shown are the adiabatic reduction to $U_1=0.04\,U_0$
 according to Eq.~(\ref{eq:uvont}), the waiting time and the ramp up.
 (b) Cumulative energy distribution: Measured fraction of the trapped
 atoms with energy below $E_0$. The horizontal axis has been scaled according
 to Fig.~\ref{fig:adiatheo}(b) using the numerical simulations to infer the
 initial atomic energy in the dipole trap. Solid Line: Fit of a cumulative
 three-dimensional Boltzmann distribution with $T = 0.09$~mK, dashed line:
 the corresponding energy distribution.}
\end{figure}

To measure the energy distribution of the atoms, we transfer them
from the MOT into the dipole trap before the trap depth is
adiabatically reduced to $U_1$ according to Eq.~(\ref{eq:uvont}).
This lowering of the potential takes between 10~ms and 51~ms for
values of $U_1$ between $0.082\,U_0$ and $0.0036\,U_0$,
respectively. After waiting for 15~ms the trap depth is ramped
back up to $U_0$ within 20~ms and the remaining atoms are
transferred back into the MOT, see Fig.~\ref{fig:energydist}(a).
The waiting time ensures that escaping atoms have travelled
sufficiently far so that they are not accidentally recaptured.

We count the initial number of atoms by observing their
fluorescence in the MOT for 50~ms before they are transferred into
the dipole trap. In the same manner we infer the number of atoms
that survived the above cooling process. We initially only load
about five atoms into the MOT to ensure that on average no more
than one atom occupies a potential well of the standing wave. For
each value of $U_1$ the above procedure was repeated 100 times to
keep the error, due to atom number statistics, below 3\%. The
change of the potential depth was realized by variation of the RF
power of the AOM drivers, while the corresponding variation of
both trap laser intensities was monitored by calibrated
photodiodes.

The result of this measurement is the cumulative energy
distribution shown in Fig.~\ref{fig:energydist}(b). Note that the
energy axis has been rescaled from the measured minimum potential
depth $U_1$ to the initial atomic energy $E_0$ using the result of
the three-dimensional trajectory simulations shown in
Fig.~\ref{fig:adiatheo}(b). Remember that in radial direction the
dipole potential is modified by gravity \cite{Schrader 01} such
that theoretically at $U_1 = 0.0031\,U_0$ the effective potential
depth is zero. It was found by extrapolation of the measured
survival probability to zero that the effective potential depth in
fact becomes zero at $U_1 = 0.0045\,U_0$, implying an actual trap
depth slightly lower than theoretically expected (see also
Sec.~\ref{sec:three-beam}). This small discrepancy has
approximately been taken into account by adding the difference of
$0.0014\,U_0$ to the theoretical values of $U_1$, which corrects
the influence of gravity for small values of $U_1$ and is
negligible at larger values.

The cumulative energy distribution of Fig.~\ref{fig:energydist}(b)
was fitted by the integral of a three-dimensional Boltzmann
distribution $p(E) \propto \sqrt{E} \exp(-E/kT)$ (shown as dashed
line). This yields a temperature of $kT=0.066\,U_0$. Using a trap
depth of $U_0=1.3 \pm 0.3$~mK we thus have $T = 0.09 \pm 0.02$~mK.
The error is due to the uncertainty in $U_0$, indicated by the
measured oscillation frequency (see Sec.\ref{sec:three-beam}).
This is slightly less than the Doppler temperature of
$T_{\text{D}}=\hbar \Gamma /2 = 0.125$~mK.

The resulting temperature of the atoms in the dipole trap is
similar to the temperatures in our high-gradient
MOT~\cite{MOT-temp}. The initial potential energy of an atom in
the dipole trap depends on its position at the time the dipole
trap is switched on. We therefore conclude that the MOT
effectively cools the atoms into the dipole trap to about $T_D$.

\section{\label{sec:three-beam}Axial oscillation frequency}

\begin{figure}
 \includegraphics[width=8.5cm]{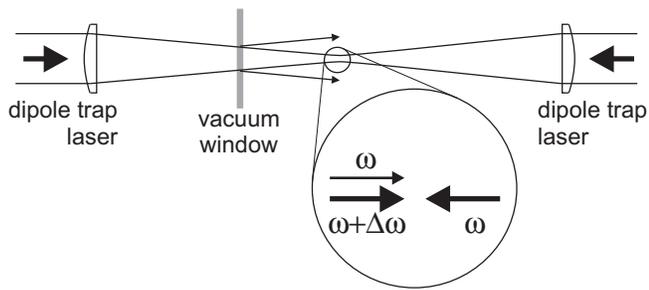}
 \caption{\label{fig:3beam}A partial
 reflection of the trapping beam at one of the vacuum cell walls interferes
 with the dipole trap.}
\end{figure}

The axial oscillation frequency $\Omega_{\text{z}}$ was measured
by resonant and parametric excitation of the oscillatory motion of
a single atom in the dipole trap, exploiting the following feature
of our experimental setup: One of the dipole trapping laser beams
passes through the window of our glass cell, which reflects about
4\% of the incident power per surface. After divergent expansion,
this third beam interferes with the two main laser beams and thus
slightly changes amplitude and phase of their interference pattern
(see Fig.~\ref{fig:3beam}). When atoms are transported by mutually
detuning the trapping beams by $\Delta \omega$ (see
Sec.~\ref{sec:dipole trap}), both phase and amplitude of the
trapping potential are modulated at that frequency. On resonance
with $\Omega_z$, this excites the oscillation of the transported
atoms, which is, in turn, used here for determining
$\Omega_{\text{z}}$.

In the atomic frame of reference moving with a velocity
$v=\lambda~\Delta\omega/4 \pi$ the total electric field is:
\begin{eqnarray}
 E(z,t) &\propto & 2  \cos(\omega t) \cos(kz)\nonumber\\  && + \beta
 \cos\bigl[ (\omega-\Delta\omega)t-k'z \bigr],
 \label{eq:efield}
\end{eqnarray}
where $\beta$ denotes the amplitude of the reflected beam in units
of the incident beam amplitude. It can be shown that the leading
terms of the resulting dipole potential for $\beta \ll 1$ and $k'
\approx k$ are given by
\begin{eqnarray}
 U(z,t) &=& U_0 \Bigl\{ \cos^2(kz) \left[ 1 + \beta \cos(\Delta\omega t)
 \right]\nonumber\\
          && -   \beta \cos(kz) \sin(kz) \sin(\Delta\omega t) \Bigr\}.
 \label{eq:3beampotential}
\end{eqnarray}
The corresponding equation of motion around the equilibrium
position $z = 0$ (assuming $kz \ll 1$) becomes:
\begin{equation}
 \ddot{z} + \Omega_{\text{z}}^2 \bigl[ 1 + \beta\cos(\Delta\omega t)\bigr] z =
 -\beta \frac{\Omega_{\text{z}}^2}{2k} \sin(\Delta\omega t).
 \label{eq:motion}
\end{equation}
It shows resonant excitation for $\Delta \omega =
\Omega_{\text{z}}$, due to the driving term on the right hand
side, as well as parametric excitation for $\Delta \omega = 2
\Omega_{\text{z}}$ due to the modulation of $\Omega_{\text{z}}$
\cite{Landau}. This leads to heating of the atoms during
transportation at mutual detunings of the laser beams near these
two values.

\begin{figure}
 \includegraphics[width=8.5cm]{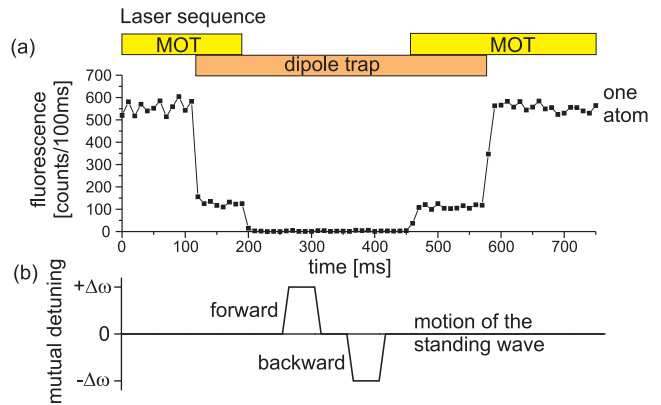}
 \caption{\label{fig:sequence}(a) Measurement procedure for the axial
 oscillation frequency. A single atom is loaded from the MOT into the
 dipole trap. During simultaneous operation of both traps, fluorescence
 of the atoms is reduced due to the light shift. Inside the dipole trap
 the atom is moved and then brought back again to the original position.
 Finally, the presence of the atom is detected by recapturing it back
 into the MOT.
 (b) Mutual detuning of the two dipole trapping beams during the transport
 (not to scale).}
\end{figure}

This resonant heating effect is used for measuring the axial
oscillation frequency $\Omega_{\text{z}}$ of the atom by keeping
$\Delta\omega$ constant for some time and by observing an increase
of the oscillation amplitude. Since the standing wave pattern of
the dipole trap moves with a velocity
$v=\lambda\Delta\omega/4\pi$, we have to accelerate and decelerate
the atom at the beginning and at the end, respectively, by
suitable short frequency ramps. Finally, the displaced atom has to
be brought back to the position of the MOT by a similar transport
in the opposite direction.

The corresponding measurement sequence is shown in
Fig.~\ref{fig:sequence}. Initially, a single atom is loaded from
the MOT into the dipole trap. The detuning $\Delta \omega$ is
ramped up quickly, then kept at a constant value to expose the
atom to the resonant heating, and finally it is ramped back down.
We limit the total transportation distance to 2~mm because further
away from the focus the trap depth, and thus $\Omega_{\text{z}}$,
decreases considerably.

\begin{figure}
 \includegraphics[width=8.5cm]{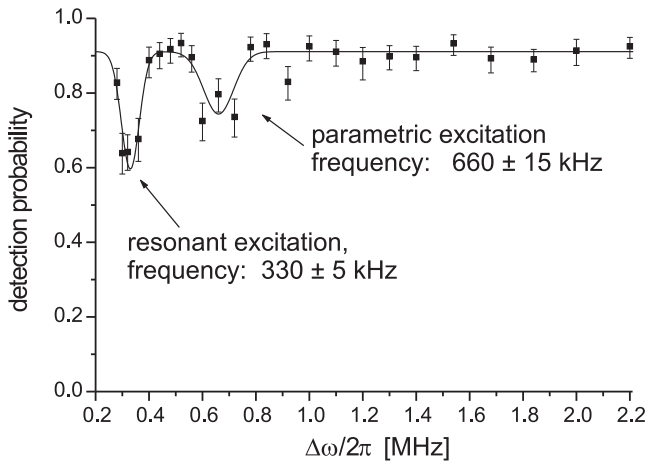}
 \caption{Measured transportation efficiency as a function of
 the atomic velocity ($v\propto\Delta\omega$). The curve is a fitted sum
 of two gaussians.}
 \label{fig:freq}
\end{figure}

Due to the anharmonicity of the trapping potential resonant
heating does not neccessarily lead to a loss of atoms. To decide
whether an atom has been resonantly heated or not, we reduce the
depth of the dipole trap in order to lose heated atoms. This is
done adiabatically, as described in Sec.~\ref{sec:adiabatic}. We
reduce the trap depth during 10~ms to 10\% of its initial value.
The reduction has been optimized to keep the atoms trapped most of
the time in the absence of resonant heating, but to lose a
substantial fraction of resonantly heated atoms. After waiting for
5~ms the potential is ramped back up and any remaining atoms are
recaptured into the MOT. The average survival probability is shown
in Fig.~\ref{fig:freq}, where we did about 100 shots with one atom
for each value of $\Delta \omega$. The clearly visible dips at
$\Delta\omega /2\pi =(330\pm 5)$~kHz and $\Delta\omega /2\pi
=(660\pm 15)$~kHz correspond to direct and parametric resonance.

The measured axial oscillation frequency agrees reasonably well
with the theoretical expectation of $\Omega_{\text{z}} / 2\pi =
380$~kHz. The discrepancy could be caused by any loss of trapping
laser intensity at the focus, e.g.~due to wavefront aberrations,
or by reduced interference contrast, e.g.~due to imperfect overlap
of the two counterpropagating beams or not perfectly matched
polarizations. Assuming 100\% interference contrast we deduce a
trap depth of $U_0 = 1.0$~mK from the measurement.

We can estimate the energy gained during the resonant excitation
as follows. During the adiabatic lowering of the trap depth to
$0.1\,U_0$ all atoms with $E_0>0.35\,U_0$ are lost
(Fig.~\ref{fig:adiatheo}(b)), leading to a survival probability of
90\% off resonance. From the cumulative energy distribution
(Fig.~\ref{fig:energydist}(b)) we see that the survival
probability of 60\% observed on resonance corresponds to a loss of
atoms with $E_0>0.1\,U_0$. These atoms must have gained an energy
of $0.25\,U_0$ during the resonant excitation period of 20~ms,
yielding a time-averaged heating rate of about 16~mK/s. In the
same way a parametric heating rate of about 13~mK/s is found.

The same resonant excitation effect considered here causes a
decrease of the transportation efficiency for certain values of
the acceleration as observed in Ref.~\cite{Schrader 01}. These
previous investigations showed that the transportation efficiency
remains nearly constant ($>95$\%) until the acceleration exceeds a
value of $10^5$~m/s$^2$. However, for certain intermediate values
of the acceleration values, at which the detuning $\Delta \omega$
matched the oscillation frequency $\Omega_{\text{z}}$, we observed
a reduction of the transportation efficiency to 75\%, which we
attribute to the resonant excitation discussed above.

\section{\label{sec:conclusions}Conclusions and outlook}

The temperature as well as the energy distribution of the atoms in
the dipole trap were measured with procedures designed to work
with single atoms. These procedures rely on our ability to
transfer single atoms between MOT and dipole trap with high
efficiency and to unambiguously detect their presence or loss. The
axial oscillation frequency was determined using controlled
transportation of the atom.

The measured temperature of 0.09~mK and oscillation frequency of
330~kHz indicate a mean oscillatory quantum number of 6. Together
with state selective detection \cite{Frese 00} this is a good
starting point for Raman cooling of a single atom to the
oscillatory ground state \cite{Hamann 98}. This will enable us to
more precisely control the internal and external degrees of
freedom of single neutral atoms.

\begin{acknowledgments}
We have received support from the Deutsche Forschungsgemeinschaft
and the state of Nordrhein-Westfalen.
\end{acknowledgments}


\begin{thebibliography}{99}


\bibitem{Grimm 00}
 R. Grimm, M. Weidem\"uller, and Y. B. Ovchinnikov,
 Adv. At. Mol. Opt. Phys. {\bf 42}, 95 (2000)

\bibitem{Balykin 00}
 V. I. Balykin, V. G. Minogin, and V. S. Letokhov,
 Rep. Prog. Phys. {\bf 63}, 1429 (2000)

\bibitem{Miller 93}
 J. D. Miller, R. A. Cline, and D. J. Heinzen,
 Phys. Rev. A {\bf 47}, R4567 (1993)

\bibitem{Davidson 95}
 N. Davidson, H. J. Lee, C. S. Adams, M. Kasevich, and S. Chu,
 Phys. Rev. Lett. {\bf 74}, 1311 (1995)

\bibitem{Milner 01}
 V. Milner, J. L. Hanssen, W. C. Campbell, and M. G. Raizen,
 Phys. Rev. Lett. {\bf 86}, 1514 (2001);
 N. Friedman, A. Kaplan, D. Carasso, and N. Davidson,
 {\it ibid.} {\bf 86}, 1518 (2001)

\bibitem{Stamper 98}
 D. M. Stamper-Kurn, M. R. Andrews, A. P. Chikkatur, S. Inouye,
 H.-J. Miesner, J. Stenger, and W. Ketterle,
 Phys. Rev. Lett. {\bf 80}, 2027 (1998);
 M. Barrett, J. Sauer, and M. S. Chapman,
 {\it ibid.} {\bf 87}, 010404 (2001);
 T. L. Gustavson, A. P. Chikkatur, A. E. Leanhardt, A. G\"{o}rlitz,
 S. Gupta, D. E. Pritchard, and W. Ketterle,
 {\it ibid.} {\bf 88}, 020401 (2002)

\bibitem{Mosk 02}
 A. Mosk, S. Kraft, M. Mudrich, K. Singer, W. Wohlleben, R. Grimm,
 and M. Weidem\"{u}ller,
 Appl. Phys. B {\bf 73}, 791 (2001)

\bibitem{Kuppens 00}
 S. J. M. Kuppens, K. L. Corwin, K. W. Miller, T. E. Chupp, and C. E. Wieman,
 Phys. Rev. A {\bf 62}, 013406 (2000)

\bibitem{Kuhr 01}
 S. Kuhr, W. Alt, D. Schrader, M. M\"uller, V. Gomer, and D. Meschede,
 Science {\bf 293}, 278 (2001), Published online 14 June 2001;
 10.1126/science.1062725

\bibitem{Schrader 01}
 D. Schrader, S. Kuhr, W. Alt, M. M\"uller, V. Gomer, and D. Meschede,
 Appl. Phys. B {\bf 73}, 819 (2001)

\bibitem{Hu 94}
 Z. Hu and H. J. Kimble,
 Opt. Lett. {\bf 19}, 1888 (1994);
 F. Ruschewitz, D. Bettermann, J. L. Peng, and W. Ertmer,
 Europhys. Lett. {\bf 34}, 651 (1996);
 D. Haubrich, H. Schadwinkel, F. Strauch, B. Ueberholz, R. Wynands,
 and D. Meschede,
 {\it ibid.} {\bf 34}, 663 (1996)

\bibitem{Frese 00}
 D. Frese, B. Ueberholz, S. Kuhr, W. Alt, D. Schrader, V. Gomer,
 and D. Meschede,
 Phys. Rev. Lett. {\bf 85}, 3777 (2000)

\bibitem{Alt 01}
 W. Alt,
 Optik {\bf 113}, 142 (2002)

\bibitem{Dalibard 85}
 J. Dalibard and C. Cohen-Tannoudji,
 J. Opt. Soc. Am. B {\bf 2}, 1707 (1985)

\bibitem{Gehm 98}
 T. A. Savard, K. M. O'Hara, and J. E. Thomas,
 Phys. Rev. A {\bf 56} R1095 (1997);
 M. E. Gehm, K. M. O'Hara, T. A. Savard, and J. E. Thomas,
 {\it ibid.} {\bf 58} 3914 (1998)

\bibitem{Landau}
 L. D. Landau and E. M. Lifschitz, {\it Mechanics} (Pergamon, New York,
 1976)

\bibitem{MOT-temp}
 A. H\"{o}pe, D. Haubrich, G. M\"{u}ller, W. G. Kaenders, and D. Meschede,
 Europhys. Lett. {\bf 22}, 669 (1993)

\bibitem{Hamann 98}
 S. E. Hamann, D. L. Haycock, G. Klose, P. H. Pax, I. H. Deutsch,
 and P. S. Jessen,
 Phys. Rev. Lett. {\bf 80}, 4149 (1998);
 H. Perrin, A. Kuhn, I. Bouchoule, and C. Salomon,
 Europhys. Lett. {\bf 42}, 395 (1998)

\end{thebibliography}
\end{document}